\documentclass[10pt]{article}

\usepackage[dvips]{graphicx}
\usepackage{eepic}
\usepackage{amsfonts}
\newcommand{\bela}[1]{\begin{equation}\label{#1}}
\newcommand{\ela}{\end{equation}}
\newcommand{\bear}[1]{\begin{array}{#1}}
\newcommand{\ear}{\end{array}}

\renewcommand{\Psi}{\mbox{\boldmath $\psi$}}

\newcommand{\as}{\\[.6em]}

\newcommand{\dis}{\displaystyle}
\renewcommand{\i}{\mbox{\rm i}}
\newcommand{\ii}{\mbox{\rm \scriptsize i}}

\newcommand{\text}{\textstyle}
\newcommand{\del}{\partial}

\newtheorem{theorem}{Theorem}

\begin{document}
\begin{center}
  \Large\bf
  Menelaus' theorem, Clifford configurations and inversive geometry of
  the Schwarzian KP hierarchy\\[8mm]
 \large\sc B.G.\ Konopelchenko\footnote{Permanent address: Dipartimento di
 Fisica, Universit\`a di Lecce and  Sezione INFN, 73100 Lecce, Italy}
  {\rm and} W.K.\ Schief\\[2mm]
  \small\sl School of Mathematics, The University of New South Wales,\\
  Sydney, NSW 2052, Australia\\[9mm]
\end{center}
\begin{abstract}
It is shown that the integrable discrete Schwarzian KP (dSKP) equation which
constitutes an algebraic superposition formula associated with, for instance,
the Schwarzian KP hierarchy, the classical Darboux transformation and 
quasi-conformal mappings  encapsulates nothing but a fundamental theorem of
ancient Greek geometry. Thus, it is demonstrated that the connection with
Menelaus' theorem and, more generally, Clifford configurations renders the dSKP
equation a natural object of inversive geometry on the plane.
The geometric and algebraic integrability of dSKP lattices and their reductions
to lattices of Menelaus-Darboux, Schwarzian KdV, Schwarzian 
Boussinesq and Schramm type is discussed.  The dSKP and discrete Schwarzian 
Boussinesq  equations are shown to represent discretizations of families of
quasi-conformal mappings.
\end{abstract}

\section{Introduction}
\setcounter{equation}{0}

The celebrated Kadomtsev-Petviashvili (KP) equation was derived some thirty 
years ago \cite{KadPet70} in connection with the 
propagation of weakly two-dimensional waves in nonlinear media. Subsequently 
\cite{ZakShab74,Dry74},
it was shown that the KP equation is amenable to the Inverse Scattering 
Transform (IST) method and hence possesses all the
remarkable properties which are
commonly associated with soliton equations (see, e.g., \cite{ZakManNovPit84,
AblCla91,JimMiw83}). In particular, there exist several infinite families of
soliton equations which are related to the KP equation such as the KP, modified
KP (mKP) and Schwarzian KP (SKP) hierarchies. Since its discovery, the KP 
equation has been the subject of extensive research. Thus, the 
KP equation and associated structures not only find application in plasma
physics and hydrodynamics but also make important appearances in various other
areas of physics and mathematics such as modern string theory 
\cite{Dij91,Mor94,FraGinZin95} and algebraic geometry~\cite{Shi86}.

In the present paper, it is demonstrated that there exist profound connections
between the KP hierarchy of modern soliton theory 
and beautiful constructions of plane geometry 
associated with the names of the Greek scholar Menelaus (Menelaos of 
Alexandria, I$^{\mbox{\scriptsize st}}$--II$^{\mbox{\scriptsize nd}}$
century) who is perhaps best known for his treatise {\em Sphaerica} on
spherical geometry and the distinguished British philosopher and mathematician
Clifford (XIX$^{\mbox{\scriptsize th}}$ century). Specifically, it is shown
that the algebraic 6-point superposition formula for the scalar SKP
hierarchy encapsulates nothing but the classical Theorem of Menelaus
and, more generally, 
Clifford's point-circle configurations. As noticed by
Clifford, the latter are naturally
embedded in the classical theory of inversive geometry and, indeed, the
6-point relation, which is known to represent a discrete Schwarzian
KP (dSKP) equation, turns out to be invariant under the group of inversive
transformations. Thus, the dSKP equation constitutes a natural object of 
inversive geometry on the plane. By construction, the SKP hierarchy represents 
an infinite set of motions of points on the plane which preserve the basic 
dSKP 6-point relation and therefore its geometric properties.

In Section 2, we demonstrate the universal nature of the dSKP 6-point relation 
by (re-)deriving it in various different settings. Contact is made with
the classical Darboux transformation and quasi-conformal mappings. In 
Section 3, the connection with Menelaus' theorem, Clifford configurations and
inversive geometry is established.
Section 4 is concerned with the interpretation of the dSKP equation as a
lattice equation. It is shown how a canonical `Lax triad' is obtained 
geometrically by introducing shape factors which encode certain angles in the
dSKP lattice. It turns out that the shape factors are governed by 
an integrable discrete KP wave function equation. If the shape factors are
real then the dSKP lattice consists of Menelaus figures. This observation
leads to a Menelaus-type reduction of the discrete Darboux system defining 
special conjugate lattices in ambient spaces of arbitrary dimension.
In Section 5, canonical dimensional reductions of the dSKP equation to
the discrete Schwarzian Korteweg-de Vries (dSKdV) and Boussinesq (dSBQ)
equations are interpreted geometrically. Degenerate Clifford 
lattices are shown to include discrete conformal mappings and 
the particular class of Schramm circle patterns.
Lattices on circles and
a Combescure-type transformation are considered. The final section deals
with the natural continuum limit of the dSKP equation and its dSKdV and
dSBQ reductions. Conformal and particular classes of quasi-conformal
mappings are obtained.

The results presented here lead to the remarkable observation that, in
principle, the complete KP theory may be retrieved from an elementary
theorem of plane geometry which is 2000 years old!

\section{The discrete Schwarzian KP equation}
\setcounter{equation}{0}

The present paper is concerned with the soliton-theoretic and geometric 
significance of the 6-point relation
\bela{E0}
  \frac{(P_1-P_2)(P_3-P_4)(P_5-P_6)}{(P_2-P_3)(P_4-P_5)(P_6-P_1)} = -1
\ela
on the complex plane. In this section, we
focus on its derivation in the context of superposition principles associated
with the Kadomtsev-Petviashvili (KP) hierarchy and quasi-conformal mappings.

\subsection{The Schwarzian KP hierarchy}

The Schwarzian KP (SKP) equation
\bela{E2.1}
  \Phi_{t_3} = \Phi_{t_1t_1t_1} + \frac{3}{2}
  \frac{\Phi_{t_2}^2-\Phi_{t_1t_1}^2}{\Phi_{t_1}} + 3W_{t_2}\Phi_{t_1},\quad 
  W_{t_1} = \frac{\Phi_{t_2}}{\Phi_{t_1}}
\ela
arises within the Painlev\'e analysis as the singularity manifold equation
associated with the KP equation \cite{Wei83}. Here, $\Phi(t_1,t_2,t_3)$ denotes
a complex-valued function and $\Phi_{t_i}=\del\Phi/\del t_i$. The SKP equation
is readily seen to be invariant under the class of M\"obius
transformations
\bela{E2.1a}
  \Phi \rightarrow \Phi' = \frac{a\Phi+b}{c\Phi+d},
\ela
where $a,b,c,d$ are arbitrary complex constants ($ad-bc\neq0$), if one 
formulates (\ref{E2.1}) in terms of the Schwarzian derivative \cite{Wei83}. It
constitutes the first member of the infinite M\"obius invariant SKP hierarchy 
of integrable equations with independent variables $t_1,t_2,t_3,t_4,\ldots$.
The latter hierarchy admits a purely algebraic superposition
formula which has been set down in \cite{BogKon98a,BogKon98b}. 
Thus, if 
\bela{E2.1b}
  \Phi = \Phi(t),\quad t = (t_1,t_2,t_3,\ldots)
\ela
constitutes a solution of the SKP hierarchy then the six solutions
\bela{E2.1c}
  \Phi_i = T_i\Phi,\quad \Phi_{ik}=T_iT_k\Phi,\quad i,k=1,2,3;\quad i\neq k,
\ela
where
\bela{E2.1d}
  T_i\Phi(t) = \Phi(t+[a_i])= \Phi 
  \left(t_1 + a_i,t_2 + \frac{a_i^2}{2},
                 t_3 + \frac{a_i^3}{3},\ldots\right)
\ela
and $a_i=\mbox{const}$, obey the 6-point relation (\ref{E0}) in the form
\bela{E2.2}
  \frac{(\Phi_1-\Phi_{12})(\Phi_2-\Phi_{23})(\Phi_3-\Phi_{13})}
  {(\Phi_{12}-\Phi_2)(\Phi_{23}-\Phi_3)(\Phi_{13}-\Phi_1)} = -1.
\ela
In the present context, the relation (\ref{E2.2}) may be interpreted as a 
lattice equation if one introduces the change of variables \cite{MIWA}
\bela{E2.2a}
  t_k = \frac{1}{k}\sum_{l=1}^{\infty}a_l^kn_l,\quad k=1,2,3,\ldots
\ela
so that the operations $T_i,\,i=1,2,3$ are associated with unit increments
of the variables $n_i$, that is, for instance,
\bela{E2.2b}
  T_1\Phi(n_1,n_2,n_3,\ldots) = \Phi(n_1+1,n_2,n_3,\ldots).
\ela
In fact, it has been shown 
\cite{BogKon98a,BogKon98b} that (\ref{E2.2}) 
constitutes a discrete version of
the SKP hierarchy and may therefore be termed discrete Schwarzian KP (dSKP) 
equation.  The dSKP equation not only encodes the SKP hierarchy but also
its B\"acklund transformation and various semi-discrete hierarchies
\cite{BogKon98a,BogKon98b,BogKon99}. It is noted that, by construction, the
dSKP equation is invariant under both the M\"obius 
transformation (\ref{E2.1a}) and the SKP flows.

The relation between $\Phi$ and the KP wave function $f$ and its dual $f^*$
is given by \cite{BogKon98a,BogKon98b,BogKon99}
\bela{E2.3}
  \Delta_i\Phi = f^*f_i,
\ela
where $\Delta_i = T_i-1$, $f_i=T_if$ and $f,f^*$ are related by
\bela{E2.4}
  \frac{f_i-f_k}{f_{ik}} = \frac{f^*_k-f^*_i}{f^*}.
\ela
Elimination of $f$ or $f^*$ gives rise to the `discrete (dual) wave function 
equations'
\bela{E2.4.5}
  \bear{rl}\dis
     \frac{f_{1}-f_{2}}{f_{12}}
   + \frac{f_{2}-f_{3}}{f_{23}}
   + \frac{f_{3}-f_{1}}{f_{13}} = & 0\\[5mm]\dis
     \frac{f_{13}^*-f_{12}^*}{f_1^*}
   + \frac{f_{12}^*-f_{23}^*}{f_2^*}
   + \frac{f_{23}^*-f_{13}^*}{f_3^*} = & 0.
  \ear
\ela
In fact, it is not difficult to 
show that the dSKP equation guarantees that there exists a parametrization of
the form (\ref{E2.3}) and the compatibility conditions
$[\Delta_i,\Delta_k]\Phi=0$ result in  (\ref{E2.4}). In addition, the 
compatibility conditions associated with the relations
\bela{E2.6}
  \frac{\Delta_i\Phi}{\Delta_k\Phi} = \frac{f_i}{f_k},\quad
  \frac{\Delta_i^*\Phi}{\Delta_k^*\Phi} 
  = \frac{f^*_{\bar{i}}}{f^*_{\bar{k}}},
\ela
where $f_{\bar{i}}=T_i^{-1}f$ and $\Delta_i^* = T_i^{-1}-1$, produce
(\ref{E2.4.5}). Conversely, if $f$ and $f^*$ are solutions of (\ref{E2.4.5})
related by (\ref{E2.4})
then the defining relations (\ref{E2.3}) for $\Phi$ are compatible and the
dSKP equation is satisfied. The algebraic superposition formulae (\ref{E2.4.5})
for the discrete (dual) KP wave functions have been derived in a different 
manner in \cite{NijCap90}.

\subsection{A superposition principle associated with the classical Darboux 
transformation}

A generic feature of integrable systems is the existence of superposition
principles (permutability theorems) associated with B\"acklund-Darboux-type 
transformations \cite{MatSal91}.
It turns out that the dSKP equation may also be derived via iterative 
application of the classical Darboux transformation \cite{Dar82} to the
`scattering problem' of the KP hierarchy
\bela{E2.2c}
  \phi_{y} = \phi_{xx} + u\phi,
\ela
where $x=t_1,y=t_2$. Thus, if $\phi$ and $\phi^i,\,i=1,2,3$ are solutions of
the Schr\"odinger equation (\ref{E2.2c}) corresponding to the potential $u$ 
then another three solutions $(\phi_i,u_i)$ are given by the Darboux transforms
\bela{E2.2d}
  \mathbb{D}_i:\qquad \phi_i = \phi_x - \frac{\phi_x^i}{\phi^i}\phi,
  \quad u_i = u + 2(\ln\phi^i)_{xx}.
\ela
The particular solutions $(\phi_i^k,u_i)$ defined by
\bela{E2.2e}
  \phi_i^k = \phi^k_x - \frac{\phi_x^i}{\phi^i}\phi^k,\quad i\neq k
\ela
may be used to obtain the `second-generation' Darboux transforms
\bela{E2.2f}
  \phi_{ik} = \phi_{ix} - \frac{\phi_{ix}^k}{\phi_i^k}\phi_i
  = \frac{\left|\bear{ccc}\phi^i&\phi_x^i&\phi_{xx}^i\\
                          \phi^k&\phi_x^k&\phi_{xx}^k\\
                          \phi&\phi_x&\phi_{xx}\ear\right|}{
    \left|\bear{cc}\phi^i&\phi_x^i\\
                   \phi^k&\phi_x^k\ear\right|}.   
\ela
The latter formulation in terms of Wronskians shows that $\phi_{ik}=\phi_{ki}$
which encapsulates the well-known permutability theorem associated with the
classical Darboux transformation:
\bela{E2.2g}
   \mathbb{D}_i\circ\mathbb{D}_k = \mathbb{D}_k\circ\mathbb{D}_i.
\ela
Due to this important commutativity property, it is possible to
interpret the action of the Darboux transformations $\mathbb{D}_i$ as shifts 
on a lattice, that is, for instance,
\bela{E2.2h}
  \phi = \phi(n_1,n_2,n_3),\quad \phi_1 = \phi(n_1+1,n_2,n_3),\quad
  \phi_{12} = \phi(n_1+1,n_2+1,n_3).
\ela
It is now readily verified that the relations (\ref{E2.2d})$_1$ and
(\ref{E2.2f}) may be combined to obtain the superposition principle (or
lattice equation)
\bela{E2.2i}
  \frac{\phi_{13}-\phi_{12}}{\phi_1} + \frac{\phi_{12}-\phi_{23}}{\phi_2}
  + \frac{\phi_{23}-\phi_{13}}{\phi_3} = 0.
\ela
The latter is precisely of the form (\ref{E2.4.5})$_2$.

In order to derive the dSKP equation, it is required to take into account the
action of the Darboux transformation on the adjoint Schr\"odinger equation
\bela{E2.2j}
   -\psi_y = \psi_{xx} + u\psi
\ela
and the associated bilinear potential $M = M(\psi,\phi)$ defined by
\bela{E2.2k}
  M_x = \psi\phi,\quad M_y = \psi\phi_x - \psi_x\phi.
\ela
Thus, it is readily verified that the pairs $(\psi_i,u_i)$, where
\bela{E2.2l}
  \psi_i = -\frac{M^i}{\phi^i},\quad M^i = M(\psi,\phi^i),
\ela
constitute solutions of the adjoint Schr\"odinger equation (\ref{E2.2j}).
Moreover, the bilinear potentials $M_i = M(\psi_i,\phi_i)$ read
\bela{E2.2m}
  M_i = M - M^i\frac{\phi}{\phi^i},
\ela
where the constants of integration have been set to zero. In particular, the
bilinear potentials $M_i^k = M(\psi_i,\phi_i^k)$ are given by
\bela{E2.2n}
  M_i^k = M^k - M^i\frac{\phi^k}{\phi^i}.
\ela
Accordingly, the second-generation bilinear potentials 
$M_{ik} = M(\psi_{ik},\phi_{ik})$ assume the form
\bela{E2.2o}
  M_{ik} = M_i - M_i^k\frac{\phi_i}{\phi_i^k}
         = M + \frac{M^i\phi^k\phi_k - M^k\phi^i\phi_i}{\phi^i\phi^k_x
             - \phi^k\phi^i_x}
\ela
so that, once again, $M_{ik}=M_{ki}$. Elimination of the quantitites 
$M_i,\phi_i$ and $\phi^i_x$ from (\ref{E2.2m}) and (\ref{E2.2o}) now
produces the superposition principle
\bela{E2.2p}
  \frac{(M_1-M_{12})(M_2-M_{23})(M_3-M_{13})}
  {(M_{12}-M_2)(M_{23}-M_3)(M_{13}-M_1)} = -1
\ela
which, regarded as a lattice equation, is nothing but the dSKP equation.

\subsection{A superposition principle for quasi-conformal mappings}

In Section 6, it is shown that, in the continuum limit, the solutions of the
dSKP equation define particular quasi-conformal mappings. The latter 
constitute mappings of the form
\bela{E2x1}
  Z : \mathbb{C} \rightarrow \mathbb{C}
\ela
defined by
\bela{E2x2}
  Z_{\bar{z}} = \mu Z_z,
\ela
where $\mu$ denotes a bounded but otherwise arbitrary function. If $\mu=0$ then
$Z_{\bar{z}}=0$ and hence $Z$ is analytic and defines a conformal 
mapping. In general, $Z$ is termed a pseudo-analytic function.
Quasi-conformal mappings have various important applications in mathematics
and physics \cite{Ahl66}-\cite{Vuo92}. In particular, they have 
been used in the study of plane gasdynamics by Bers \cite{Ber58}.

The Beltrami equation (\ref{E2x2}) is invariant under the Darboux-type
transformations
\bela{E2x3}
  \mathbb{D}_i:\qquad 
  Z \rightarrow Z_i = Z - \frac{Z^i}{Z^i_z}Z_z,
\ela
where $Z^i,\,i=1,2,3$ represent another three solutions of the Beltrami
equation with the same potential $\mu$. It is noted that $\mu$ is preserved
by $\mathbb{D}_i$. As in Section~2.2, the particular solutions
\bela{E2x4}
  Z_i^k = Z^k - \frac{Z^i}{Z^i_z}Z^k_z,\quad i\neq k
\ela
generate the Darboux transforms
\bela{E2x5}
  Z_{ik} = Z_i - \frac{Z_i^k}{Z^k_{iz}}Z_{iz}.
\ela
Thus, the commutativity property 
$Z_{ik}=Z_{ki}$ holds and, as a consequence, the six relations 
(\ref{E2x5}) may be combined to obtain 
\bela{E2x6}
  \frac{(Z_1-Z_{12})(Z_2-Z_{23})(Z_3-Z_{13})}
  {(Z_{12}-Z_2)(Z_{23}-Z_3)(Z_{13}-Z_1)} = -1.
\ela
Accordingly, the dSKP equation may also be regarded as a superposition
principle for quasi-conformal mappings or pseudo-analytic functions.

\subsection{Tau-function parametrizations}

It is well-known (see, e.g., \cite{JimMiw83}) that the KP hierarchy and
associated quantities may be expressed in terms of tau-functions which obey
the discrete Hirota equation
\bela{E2.6a}
  \tau_{12}\tau_3 = \tau_{23}\tau_1 + \tau_{13}\tau_2.
\ela
For instance, the dual wave function $f^*$ is related to a tau-function $\tau$
via
\bela{E2.11}
  f^*_2 = f^*_1 + \frac{\tau_{12}\tau}{\tau_1\tau_2}f^*,\quad
  f^*_3 = f^*_1 + \frac{\tau_{13}\tau}{\tau_1\tau_3}f^*.
\ela
This linear system is compatible modulo the discrete Hirota equation and 
the discrete dual wave function equation (\ref{E2.4.5})$_2$ is identically 
satisfied. 

Now, on the one hand, the quantity
\bela{E2.7a}
  \tilde{\tau} = f^*\tau
\ela
is readily shown to be another solution of the discrete Hirota equation. In 
fact, $\tilde{\tau}$ constitutes a discrete Darboux transform of $\tau$.
Thus, the dual KP wave function admits the representation
\bela{E2.7}
  f^* = \frac{\tilde{\tau}}{\tau},
\ela
where $\tilde{\tau}$ is a solution of the compatible system
\bela{E2.8}
  \bear{rl}
   \tilde{\tau}_2\tau_1 = & \tilde{\tau}_1\tau_2 + \tilde{\tau}\tau_{12}\as
   \tilde{\tau}_3\tau_1 = & \tilde{\tau}_1\tau_3 + \tilde{\tau}\tau_{13}\as
   \tilde{\tau}_2\tau_3 = & \tilde{\tau}_3\tau_2 + \tilde{\tau}\tau_{23}.
  \ear
\ela
The equations (\ref{E2.8})$_{1,2}$ are obtained by inserting $f^*$ as given by
(\ref{E2.7}) into the linear system (\ref{E2.11}). The remaining relation 
(\ref{E2.8})$_3$ is redundant and has been added merely for reasons of 
symmetry.  

On the other hand, the quantity 
\bela{E2.11a}
  \hat{\tau} = \Phi\tau
\ela
also constitutes another solution of the discrete Hirota equation. This 
solution may be interpreted as a discrete binary Darboux transform of the
seed solution~$\tau$. Accordingly, any solution of the dSKP
equation may be written as the ratio of two tau-functions, namely
\bela{E2.11b}
  \Phi = \frac{\hat{\tau}}{\tau}.
\ela

In conclusion, it is noted that the integrable nature of the dSKP 
equation (\ref{E2.2}) and the discrete (dual) KP wave function equations 
(\ref{E2.4.5})
is inherited from the integrable (Schwarzian) KP hierarchy. In particular,
the IST and \mbox{$\bar{\del}$-dressing} methods and B\"acklund transformations 
may be employed to construct large classes of corresponding solutions 
\cite{Dry74,ZakManNovPit84,AblCla91,MatSal91}. On the other hand, the fact
that the dSKP equation resembles some equations of integrable discrete 
geometry \cite{BobSei99} and is invariant under M\"obius transformations 
suggests that the dSKP equation may be interpreted geometrically. Such belief
has been expressed in \cite{BogKon99} but no satisfactory geometric 
interpretation has been provided. Here, we address this problem and place the 
dSKP equation in a rather unexpected geometric setting.

\section{Menelaus' theorem, Clifford 
  configurations and the local dSKP equation}
\setcounter{equation}{0}

In the preceding, it has been shown that the 6-point relation (\ref{E0}) 
may be identified with the {\em local} dSKP equation (\ref{E2.2}). 
Now, the remarkable
observation is that this 6-point relation encapsulates nothing but a
theorem which was set down some 2000 years ago and `is of great importance' 
\cite{Ask77} in plane geometry (see, e.g., \cite{Ped70,BraEspGra00}). It
bears the name of Menelaus and may be regarded as an axiom of affine
geometry \cite{axiom}. 

\subsection{Menelaus' theorem}

\begin{figure}[h]
\begin{center}
\setlength{\unitlength}{0.00037489in}
\begin{picture}(12000,4500)(500,1108)
\put(1282,1468){\blacken\ellipse{100}{100}}
\put(3982,1468){\blacken\ellipse{100}{100}}
\put(5332,1468){\blacken\ellipse{100}{100}}
\put(4432,3268){\blacken\ellipse{100}{100}}
\put(4882,5068){\blacken\ellipse{100}{100}}
\put(3982,4168){\blacken\ellipse{100}{100}}
\path(1282,1468)(4882,5068)
\path(4882,5068)(3982,1468)
\path(3982,4168)(5332,1468)
\path(5332,1468)(1282,1468)
\put(1192,1138){$A$}
\put(3892,1108){$B$}
\put(4837,5278){$C$}
\put(5407,1138){$F$}
\put(3737,4303){$E$}
\put(4597,3243){$D$}
\put(1192,3243){$\mathbb{R}^2$}
%
\put(7282,1468){\blacken\ellipse{100}{100}}
\put(9982,1468){\blacken\ellipse{100}{100}}
\put(11332,1468){\blacken\ellipse{100}{100}}
\put(10432,3268){\blacken\ellipse{100}{100}}
\put(10882,5068){\blacken\ellipse{100}{100}}
\put(9982,4168){\blacken\ellipse{100}{100}}
\path(7282,1468)(10882,5068)
\path(10882,5068)(9982,1468)
\path(9982,4168)(11332,1468)
\path(11332,1468)(7282,1468)
\put(7192,1138){$\Phi_1$}
\put(9892,1108){$\Phi_2$}
\put(10837,5278){$\Phi_3$}
\put(11407,1138){$\Phi_{12}$}
\put(9287,4303){$\Phi_{13}$}
\put(10597,3243){$\Phi_{23}$}
\put(7192,3243){$\mathbb{C}$}
\end{picture}
\end{center}
\caption{A Menelaus figure}
\label{menelaus}
\end{figure}
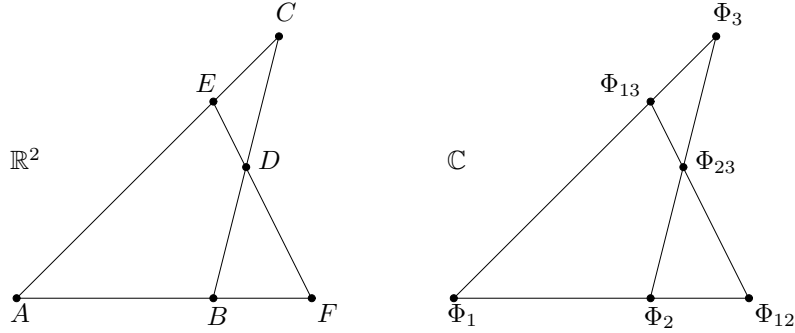

\begin{theorem} {\bf (Menelaus' theorem [I$^{\mbox{\scriptsize st}}$ 
century AD] and its converse).}
Let $A,B,C$ be the vertices of a triangle and $D,E,F$ be three points on
the (extended) edges of the triangle opposite to $A,B,C$ respectively 
(see Figure~\ref{menelaus}). Then, the points $D,E,F$ are collinear if and only
if
\bela{E3.1}
  \frac{\overline{AF}}{\overline{FB}}\, 
  \frac{\overline{BD}}{\overline{DC}}\,
  \frac{\overline{CE}}{\overline{EA}} 
  = -1,
\ela
where $\overline{PQ}/\overline{QR}$ denotes the ratio of directed lengths
associated with any three collinear points $P,Q,R$.
\end{theorem}

The key relation (\ref{E3.1}) is indeed of the form (\ref{E0}) or (\ref{E2.2})
but it is formulated in terms of real numbers. Its complex-valued version
(see, e.g.,~\cite{Les96}) is obtained by regarding the plane spanned by the 
triangle as the complex plane and labelling the points $A,\ldots,F$ by complex
numbers according to (see Figure~\ref{menelaus})
\bela{E3.1a}
  A = \Phi_1,\quad B=\Phi_2,\quad C=\Phi_3,\quad D = \Phi_{23},\quad
  E = \Phi_{13},\quad F = \Phi_{12}.
\ela
Due to the collinearity of the points $(A,B,F)$, $(B,D,C)$  and $(A,E,C)$, the
ratios
\bela{E3.3}
  \alpha = \frac{\Phi_{12}-\Phi_1}{\Phi_{12}-\Phi_2},\quad
  \beta  = \frac{\Phi_{23}-\Phi_2}{\Phi_{23}-\Phi_3},\quad
  \gamma = \frac{\Phi_{13}-\Phi_3}{\Phi_{13}-\Phi_1}
\ela
are real and, indeed,
\bela{E3.3a}
  \alpha = - \frac{\overline{AF}}{\overline{FB}},\quad
  \beta  = - \frac{\overline{BD}}{\overline{DC}},\quad
  \gamma = - \frac{\overline{CE}}{\overline{EA}}.
\ela
Thus, if we define the multi-ratio
\bela{E3.3b}
  M(P_1,P_2,P_3,P_4,P_5,P_6) = 
  \frac{(P_1-P_2)(P_3-P_4)(P_5-P_6)}{(P_2-P_3)(P_4-P_5)(P_6-P_1)}
\ela
then the Menelaus relation (\ref{E3.1}) may be cast into the form
\bela{E3.2}
 M(\Phi_1,\Phi_{12},\Phi_2,\Phi_{23},\Phi_3,\Phi_{13}) = -1.
\ela
We therefore conclude that any six points associated with a `Menelaus figure'
(Figure~\ref{menelaus}) give rise to a particular solution of the local dSKP 
equation.  

The general solution of (\ref{E3.2}) is obtained by exploiting its symmetry
group. It is evident that, in addition to the M\"obius transformation 
(\ref{E2.1a}), equation (\ref{E3.2}) is invariant under complex conjugation
$\Phi\rightarrow\bar{\Phi}$. The set of these transformations form the group
of inversive transformations on the plane. It consists of
Euclidean motions, scalings and inversions in circles. Analytically, the
latter are represented~by transformations of the form
\bela{E3.4}
  \Phi\rightarrow\Phi' = \Phi_* + \frac{r^2}{\bar{\Phi}-\bar{\Phi}_*},
\ela
where $r$ and $\Phi_*$ 
denote the radius and centre of a circle respectively. The 
inversion (\ref{E3.4}) converts circles passing through the point $\Phi_*$ into
straight lines and vice versa while generic circles are mapped to circles.
Inversive geometry is concerned with the properties of figures on the plane
which are preserved by inversive transformations \cite{BraEspGra00,MorMor54}.
Accordingly, the 6-point relation (\ref{E3.2}) and therefore the dSKP
equation are objects of inversive geometry on the plane. 

\begin{figure}[h]
\begin{center}
\setlength{\unitlength}{0.00067489in}
\begin{picture}(5177,3837)(0,-10)
\put(2194,2329){\ellipse{1972}{1972}}
\put(867,1864){\ellipse{1718}{1718}}
\put(1392,2277){\ellipse{1626}{1626}}
\put(3399,1777){\ellipse{3540}{3540}}
\put(712,2704){\blacken\ellipse{80}{80}}
\put(1262,2614){\blacken\ellipse{80}{80}}
\put(1682,2149){\blacken\ellipse{80}{80}}
\put(1982,2824){\blacken\ellipse{80}{80}}
\put(2462,3264){\blacken\ellipse{80}{80}}
\put(1557,3069){\blacken\ellipse{80}{80}}
\put(1652,1509){\shade\ellipse{80}{80}}
\put(1732,2007){$\Phi_1$}
\put(2107,2744){$\Phi_{13}$}
\put(2252,3379){$\Phi_3$}
\put(337,2754){$\Phi_{12}$}
\put(1292,3212){$\Phi_{23}$}
\put(1377,2549){$\Phi_2$}
\put(1502,1159){$\Phi_*$}
\end{picture}
\end{center}
\caption{A Menelaus configuration}
\label{clifford}
\end{figure}
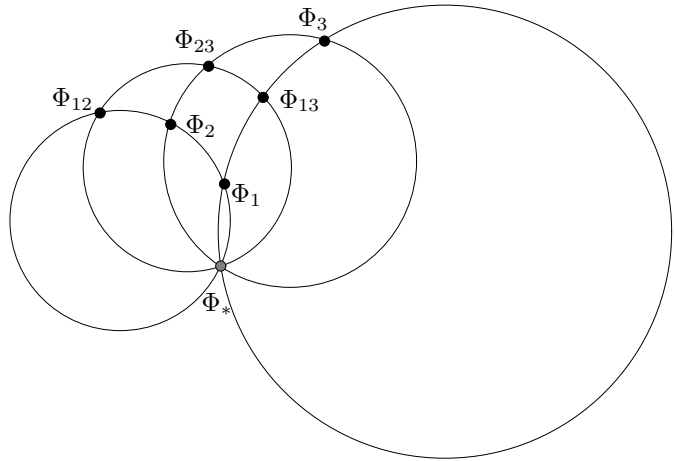

If we now apply a generic inversion to a Menelaus figure (Figure 
\ref{menelaus}) then we obtain four circles passing through the point 
$\Phi_*$ as shown in Figure \ref{clifford}.
It is natural to term this figure `Menelaus configuration' since it may be
mapped to a Menelaus figure by means of an inversion with
respect to the point of intersection~$\Phi_*$ without changing the
multi-ratio $M(\Phi_1,\Phi_{12},\Phi_2,\Phi_{23},\Phi_3,\Phi_{13})$. A simple 
counting argument indicates that the set of Menelaus configurations should 
encapsulate the general solution of
(\ref{E3.2}). Indeed, the following theorem which constitutes a geometric
interpretation of the 6-point relation (\ref{E3.2}) holds:

\begin{theorem} {\bf (Inversive geometry of the local dSKP equation).}
Let $\Phi_1,\Phi_{12},\Phi_{2},\Phi_{23},\Phi_{3},\Phi_{13}$ be
six generic points on the complex plane. Then, the four circles passing 
through $(\Phi_1,\Phi_{12},\Phi_2)$, $(\Phi_2,\Phi_{23},\Phi_3)$, 
$(\Phi_3,\Phi_{13},\Phi_1)$ and $(\Phi_{12},\Phi_{23},\Phi_{13})$ meet at 
a point $\Phi_*$ if and only if
\bela{E3.4a}
  M(\Phi_1,\Phi_{12},\Phi_2,\Phi_{23},\Phi_3,\Phi_{13}) = -1.
\ela
\end{theorem}

\noindent
{\bf Proof.} On the one hand, if the four circles intersect at a point $\Phi_*$
then one may apply an inversion with respect to that point and a Menelaus 
figure is obtained. According to the Menelaus theorem, the associated 
multi-ratio is equal to $-1$ and since the multi-ratio is invariant 
under inversive transformations (up to complex conjugation), we deduce that 
(\ref{E3.4a}) holds.

On the other hand, let us assume that the 6-point relation (\ref{E3.4a}) holds.
We first draw two circles passing through the points 
$(\Phi_1,\Phi_{12},\Phi_2)$ and $(\Phi_3,\Phi_{13},\Phi_1)$. Since we 
consider the generic situation, these intersect at $\Phi_1$ and some 
point~$\Phi_*$, say. Another two circles are now determined by the triplets
$(\Phi_2,\Phi_*,\Phi_3)$ and $(\Phi_{12},\Phi_*,\Phi_{13})$. The latter pair
of circles intersect at $\Phi_*$ and some point $\Phi_{\circ\circ}$. 
The argument 
employed in the first part of this proof now shows that the points
$\Phi_1,\Phi_{12},\Phi_{2},\Phi_{\circ\circ},\Phi_{3},\Phi_{13}$ and the 
corresponding four circles which meet at $\Phi_*$ form a Menelaus 
configuration, whence
\bela{E3.4b}
  M(\Phi_1,\Phi_{12},\Phi_2,\Phi_{\circ\circ},\Phi_3,\Phi_{13}) = -1.
\ela
However, since both $\Phi_{23}$ and $\Phi_{\circ\circ}$ are solutions 
of the same linear equation (\ref{E3.4a}) (or (\ref{E3.4b})), it is concluded 
that $\Phi_{\circ\circ}$ coincides with $\Phi_{23}$ and the proof is complete.
\medskip

\subsection{Clifford configurations}

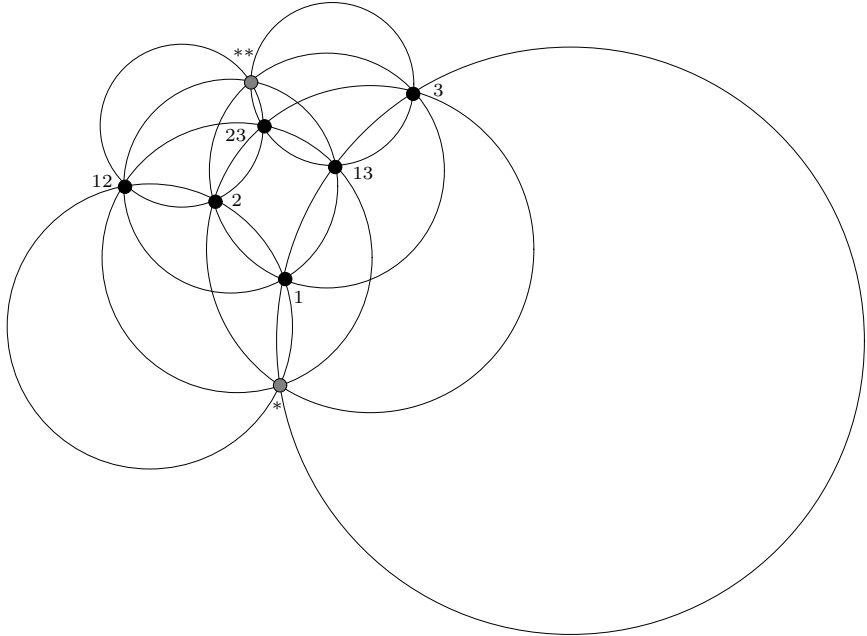
\begin{figure}[h]
\begin{center}
\setlength{\unitlength}{0.00087489in}
\begin{picture}(5177,3837)(0,-10)
\put(2194,2329){\ellipse{1972}{1972}}
\put(867,1864){\ellipse{1718}{1718}}
\put(1392,2277){\ellipse{1626}{1626}}
\put(3399,1777){\ellipse{3540}{3540}}
\put(1354,2708){\ellipse{1290}{1290}}
\put(1965,3323){\ellipse{984}{984}}
\put(1058,3072){\ellipse{984}{984}}
\put(1932,2802){\ellipse{1416}{1416}}
\put(717,2704){\blacken\ellipse{80}{80}}
\put(1262,2614){\blacken\ellipse{80}{80}}
\put(1682,2149){\blacken\ellipse{80}{80}}
\put(1982,2824){\blacken\ellipse{80}{80}}
\put(2452,3264){\blacken\ellipse{80}{80}}
\put(1557,3069){\blacken\ellipse{80}{80}}
\put(1652,1509){\shade\ellipse{80}{80}}
\put(1477,3334){\shade\ellipse{80}{80}}
\put(1732,2007){$\scriptstyle1$}
\put(2087,2754){$\scriptstyle13$}
\put(2572,3249){$\scriptstyle3$}
\put(517,2704){$\scriptstyle12$}
\put(1322,2982){$\scriptstyle23$}
\put(1357,2589){$\scriptstyle2$}
\put(1602,1359){$\scriptstyle*$}
\put(1367,3489){$\scriptstyle**$}
\end{picture}
\end{center}
\caption{A $\mathcal{C}_4$ Clifford configuration}
\label{clifford4}
\end{figure}

It is evident that if (\ref{E3.4a}) holds then
\bela{E3.4c}
  M(\Phi_{12},\Phi_2,\Phi_{23},\Phi_3,\Phi_{13},\Phi_1) = -1.
\ela
This implies, in turn, that the four circles passing through 
$(\Phi_{13},\Phi_1,\Phi_{12})$,
\linebreak
$(\Phi_{12},\Phi_2,\Phi_{23})$, 
$(\Phi_{23},\Phi_3,\Phi_{13})$ and $(\Phi_1,\Phi_2,\Phi_3)$ likewise meet
at a point $\Phi_{**}$ as illustrated in Figure \ref{clifford4}.
Inspection of Figure \ref{clifford4} shows that the eight circles and 
eight points are positioned in such a way that
there exist four circles passing through each point and four points lying on
each circle. Configurations of this kind are known as $\mathcal{C}_4$
Clifford configurations \cite{Zie41,Lon76}. These are constructed in the
following way. Consider a point $P$ on the plane and four generic circles
$S_1,S_2,S_3,S_4$ passing through $P$. The additional six points of 
intersection are labelled by $P_{12},P_{13},P_{14},P_{23},P_{24},P_{34}$. 
Here, the indices on $P_{ik}$ correspond to those of the circles
$S_i$ and $S_k$. By
virtue of Theorem 2, the associated multi-ratio is given by
\bela{E3.4d}
  M(P_{14},P_{12},P_{24},P_{23},P_{34},P_{13}) = -1.
\ela
Any three circles $S_i,S_k,S_l$ intersect at three points and therefore define
a circle $S_{ikl}$ passing through these points. Clifford's circle theorem
\cite{Cli74} (and Theorem~2) then states that the four circles 
$S_{123},S_{124},S_{134},S_{234}$ meet at a point $P_{1234}$. Thus, Figure
\ref{clifford4} indeed displays a $\mathcal{C}_4$ configuration with, for 
instance,
\bela{E3.4.e}
 \bear{rlrlrlrl}
  P = &\Phi_{**},&\quad P_{12} = &\Phi_{12},&\quad P_{13} = &\Phi_{13},&\quad
  P_{14} = &\Phi_1\as
  P_{23} = &\Phi_{23},&\quad P_{24} = &\Phi_2,&\quad
  P_{34} = &\Phi_3,&\quad P_{1234} = &\Phi_{*}.
 \ear
\ela
Remarkably, Clifford configurations ($\mathcal{C}_n$) exist for any number of
initial circles $S_1,\ldots,S_n$ passing through a point $P$. 

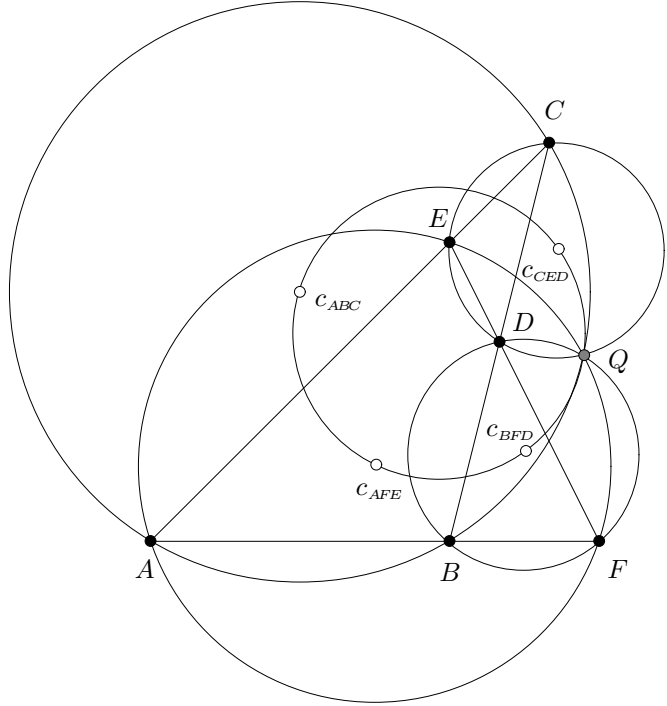
\begin{figure}[h]
\begin{center}
\setlength{\unitlength}{0.00057489in}
\begin{picture}(5929,6600)(0,100)
\put(2632,3718){\ellipse{5248}{5248}}
\put(3307,2143){\ellipse{4270}{4270}}
\put(4949,4093){\ellipse{1944}{1944}}
\put(1282,1468){\blacken\ellipse{100}{100}}
\put(3982,1468){\blacken\ellipse{100}{100}}
\put(5332,1468){\blacken\ellipse{100}{100}}
\put(4432,3268){\blacken\ellipse{100}{100}}
\put(4882,5068){\blacken\ellipse{100}{100}}
\put(3982,4168){\blacken\ellipse{100}{100}}
\put(4649,2248){\ellipse{2086}{2086}}
\put(3888,3344){\ellipse{2642}{2642}}
\put(2632,3718){\whiten\ellipse{100}{100}}
\put(3322,2158){\whiten\ellipse{100}{100}}
\put(4967,4108){\whiten\ellipse{100}{100}}
\put(4672,2283){\whiten\ellipse{100}{100}}
\put(5197,3148){\shade\ellipse{100}{100}}
\path(1282,1468)(4882,5068)
\path(4882,5068)(3982,1468)
\path(3982,4168)(5332,1468)
\path(5332,1468)(1282,1468)
\put(1142,1138){$A$}
\put(3892,1108){$B$}
\put(4837,5278){$C$}
\put(5407,1138){$F$}
\put(3787,4303){$E$}
\put(4547,3363){$D$}
\put(3137,1878){$c_{\scriptscriptstyle A\!F\!E}$}
\put(4312,2413){$c_{\scriptscriptstyle B\!F\!D}$}
\put(4632,3853){$c_{\scriptscriptstyle C\!E\!D}$}
\put(2767,3593){$c_{\scriptscriptstyle A\!B\!C}$}
\put(5412,3028){$Q$}
\end{picture}
\end{center}
\caption{A Wallace configuration}
\label{wallace}
\end{figure}

The circles and points of a ($\mathcal{C}_4$) Clifford configuration 
are known to appear on equal footing in the sense that the
angles made by the four oriented circles passing through a point
are the same for all eight points \cite{Zie41}. Indeed, if we apply an 
inversion with 
respect to any of the eight points $P,\ldots,P_{1234}$ then four circles are 
mapped to four 
straight lines and the images of the remaining four circles meet at a point.
Thus, we obtain a Menelaus figure and retrieve Wallace's theorem~\cite{Wal06}
which states that the four circumcircles of the triangles formed by four lines
pass through a point. This is illustrated in Figure~\ref{wallace}.
Due to the symmetry of the $\mathcal{C}_4$ configuration, 
corresponding lines of the eight Menelaus 
figures which are obtained by inversion and appropriate rotation are parallel.

It is noted in passing that there exists an elegant way of parametrizing
Clifford and Menelaus configurations \cite{MorMor54}. This may be exploited
to prove in a purely algebraic manner, for instance, Steiner's theorem 
\cite{Ste27} which states that in a Wallace
configuration the centres and the point of intersection of the four 
circumcircles lie on a circle (cf.~Figure \ref{wallace}).

\section{Clifford and Menelaus lattices. Geometric integrability}
\setcounter{equation}{0}

In the previous section, the connection between the local dSKP equation and
Menelaus configurations has been established. We now return to the 
interpretation of the dSKP equation
\bela{E4.1}
  M(\Phi_1,\Phi_{12},\Phi_2,\Phi_{23},\Phi_3,\Phi_{13}) = -1
\ela
as a lattice equation for maps of the form
\bela{E4.2}
   \Phi : \mathbb{Z}^3\rightarrow \mathbb{C}.
\ela
As in Section 2, the indices on $\Phi$ are interpreted as shifts on the 
lattice. Accordingly, lattices of the form (\ref{E4.2}) subject to the
dSKP equation (\ref{E4.1}) are integrable and consist of an infinite number of
Menelaus configurations. In order to distinguish these lattices from lattices
which are composed of Menelaus figures, we refer to these lattices as
Clifford lattices. In this connection, it is noted that any Menelaus 
configuration of a Clifford lattice may be mapped to a Menelaus figure by
means of an inversion. However, in general, the remaining Menelaus 
configurations are not converted into Menelaus figures. We here embark on a
study of the `geometric integrability' of Clifford lattices and show that
Menelaus lattices indeed exist and are likewise integrable. The latter turn out
to be related to a canonical reduction of the integrable discrete Darboux
system \cite{BogKon95,Dol97}.

\subsection{Clifford lattices}

We begin with a lattice $\Phi$ of the form (\ref{E4.2}) which contains one
Menelaus configuration, that is there exist six points 
$\Phi_1,\Phi_{12},\Phi_2,\Phi_{23},\Phi_3,\Phi_{13}$ such that (\ref{E4.1})
holds. It is natural to introduce complex parameters $\alpha,\beta$ and 
$\gamma$ according to (cf.\ (\ref{E3.3}))
\bela{E4.3}
  \alpha = \frac{\Phi_{12}-\Phi_1}{\Phi_{12}-\Phi_2},\quad
  \beta  = \frac{\Phi_{23}-\Phi_2}{\Phi_{23}-\Phi_3},\quad
  \gamma = \frac{\Phi_{13}-\Phi_3}{\Phi_{13}-\Phi_1}.
\ela
These encode the angles made by the edges of the triangles 
$\Delta(\Phi_1,\Phi_{12},\Phi_2)$, $\Delta(\Phi_2,\Phi_{23},\Phi_3)$ and
$\Delta(\Phi_3,\Phi_{13},\Phi_1)$ respectively and therefore the equivalence
classes of triangles which are similar \cite{Les96}. The definitions 
(\ref{E4.3}) may be
regarded as linear equations for $\Phi_1,\Phi_{12},\Phi_2,\Phi_{23},\Phi_3,
\Phi_{13}$ with the `shape' parameters $\alpha,\beta,\gamma$ as coefficients:
\bela{E4.4}
 \bear{rl}
  \Phi_{12}-\Phi_1 = &\alpha(\Phi_{12}-\Phi_2)\as
  \Phi_{23}-\Phi_2 = &\beta(\Phi_{23}-\Phi_3)\as
  \Phi_{13}-\Phi_3 = &\gamma(\Phi_{13}-\Phi_1).
 \ear
\ela
The Menelaus configuration is then encapsulated in the relation
\bela{E4.5}
  \alpha\beta\gamma = 1.
\ela
Shape parameters $\alpha_3,\beta_1,\gamma_2$ are associated with the
triangles $\Delta(\Phi_{13},\Phi_{123},\Phi_{23})$, 
$\Delta(\Phi_{12},\Phi_{123},\Phi_{13})$ and
$\Delta(\Phi_{23},\Phi_{123},\Phi_{12})$ respectively so that
\bela{E4.6}
  \bear{rl}
  \Phi_{123}-\Phi_{13} = &\alpha_3(\Phi_{123}-\Phi_{23})\as
  \Phi_{123}-\Phi_{12} = &\beta_1(\Phi_{123}-\Phi_{13})\as
  \Phi_{123}-\Phi_{23} = &\gamma_2(\Phi_{123}-\Phi_{12})
 \ear
\ela
which immediately implies that
\bela{E4.7}
  \alpha_3\beta_1\gamma_2 = 1.
\ela
The remaining consistency condition is obtained by eliminating $\Phi_{123}$
from any two of the relations (\ref{E4.6}). It turns out that, remarkably,
elimination of $\Phi_{123}$ from (\ref{E4.6})$_{1,2}$ leads to a constraint
on the shape parameters only, namely
\bela{E4.8}
  (\alpha_3\beta_1-1)(\alpha-1) = (\alpha\beta-1)(\alpha_3-1).
\ela
It is emphasized that the above relation is invariant under cyclic permutation.
Thus, the following theorem may be formulated:

\begin{theorem} {\bf (Clifford and Menelaus lattices).} A lattice $\Phi : 
\mathbb{Z}^3\rightarrow\mathbb{C}$ is composed of Menelaus configurations 
or figures if
and only if the associated complex or real shape parameters, regarded as 
functions, satisfy the lattice equations
\bela{E4.9}
  \alpha\beta\gamma = 1,\quad \alpha_3\beta_1\gamma_2 = 1,\quad 
   (\alpha_3\beta_1-1)(\alpha-1) = (\alpha\beta-1)(\alpha_3-1).
\ela
\end{theorem}

{}From the point of view of integrable systems, it is natural to investigate
the compatibility of the linear system (\ref{E4.4}) without referring to the
genesis of the shape parameters. Thus, the compatibility conditions 
$(\Phi_{12})_3 = (\Phi_{23})_1 = (\Phi_{13})_2$ lead to two equations of the
form
\bela{E4.10}
  E^{1k}(\Phi_1-\Phi_2) + E^{2k}(\Phi_2-\Phi_3) = 0,\quad k=1,2,
\ela
where the coefficients $E^{ik}$ depend on the shape parameters. It turns out
that these coefficients vanish if and only if the shape parameters satisfy
(\ref{E4.9}). Hence, the `geometric integrability' of Clifford lattices 
coincides with their
algebraic integrability and, in fact, (\ref{E4.4}) is nothing but a {\em
linear} triad representation of the {\em nonlinear} system (\ref{E4.9}) for 
the shape parameters.
Moreover, the relations (\ref{E4.9})$_{1,2}$ imply the
parametrization
\bela{E4.11}
   \alpha = \frac{\phi_1}{\phi_2},\quad \beta = \frac{\phi_2}{\phi_3},\quad
   \gamma = \frac{\phi_3}{\phi_1}
\ela
in terms of some function $\phi$. The remaining relation (\ref{E4.9})$_3$
then reads
\bela{E4.12}
  \frac{\phi_{13}-\phi_{12}}{\phi_1} + \frac{\phi_{12}-\phi_{23}}{\phi_2}
  + \frac{\phi_{23}-\phi_{13}}{\phi_3} = 0
\ela
which constitutes the integrable discrete dual KP wave function equation
(\ref{E2.4.5})$_2$ with $\phi = f^*$. Accordingly, the shape parameters of
Clifford lattices admit the eigenfunction and tau-function representations
\bela{E4.13}
  \alpha = \frac{f^*_1}{f^*_2} = 
  \frac{\tilde{\tau}_1\tau_2}{\tilde{\tau}_2\tau_1},\quad
  \beta = \frac{f^*_2}{f^*_3} =
  \frac{\tilde{\tau}_2\tau_3}{\tilde{\tau}_3\tau_2},\quad
  \gamma = \frac{f^*_3}{f^*_1} =
  \frac{\tilde{\tau}_3\tau_1}{\tilde{\tau}_1\tau_3}.
\ela

It is emphasized that even though we have adopted a particular
parametrization of the Menelaus figures and configurations in the construction
of associated lattices, the symmetry
of the Menelaus figures and configurations is still present in the sense that
all lines and circles appear on equal footing. In fact, if we regard any three
collinear points of a Menelaus figure as the vertices of a degenerate triangle
then Menelaus lattices, that is lattices which are composed of Menelaus 
figures, admit the combinatorics of face-centred cubic (fcc)
lattices. Indeed, if we use the reparametrization
\bela{I1}
  n_4 = n_2+n_3-1,\quad n_5=n_1+n_3-1,\quad n_6=n_1+n_2-1
\ela
then the lattice $\Phi$ assumes the form
\bela{I2}
  \bear{rl}
  \Phi : G&\rightarrow\mathbb{C}\as
         G&=\{(n_4,n_5,n_6)\in\mathbb{Z}^3:n_4+n_5+n_6\,\mbox{ odd}\}.
  \ear
\ela
The edge structure of $G$ induced by the edges of the
`octahedral' Menelaus figures is
obtained by starting at the vertex $(1,0,0)$ and drawing diagonals across the
faces of the cubic lattice $\mathbb{Z}^3$. Thus, $G$ is composed of octahedra
and tetrahedra and constitutes an fcc lattice with
$\Phi_{1},\Phi_{12},\Phi_{2},\Phi_{23},\Phi_{3},\Phi_{13}$ or, equivalently,
$\Phi_{\bar{4}},\Phi_6,\Phi_{\bar{5}},\Phi_4,\Phi_{\bar{6}},\Phi_5$ being
the vertices of the octahedra. Accordingly, the Menelaus figures (and 
configurations) may be regarded as the images of the octahedra under the 
mapping $\Phi$. 

\subsection{The Menelaus reduction of Darboux lattices}

In the preceding, it has been demonstrated that the fact that the Menelaus 
configuration can be extended to a Clifford lattice encodes its integrability.
Here, we investigate this observation in more detail and show that the 
extensibility of a figure to a lattice may constitute a `test' for 
integrability. Thus, for simplicity, we confine ourselves to
lattices with real shape parameters. Examples of such lattices are Menelaus
lattices which correspond
to real solutions of the shape equations (\ref{E4.9}) or, equivalently, the
discrete dual KP wave equation~(\ref{E4.12}). Indeed, for any such solution,
the linear system  (\ref{E4.4}) is compatible and
$M(\Phi_1,\Phi_{12},\Phi_2,\Phi_{23},\Phi_3,\Phi_{13}) = -1$. 

By virtue of the Ceva theorem given below, the linear system
(\ref{E4.4}) may likewise be associated with classical Ceva figures
\cite{Ped70,BraEspGra00} as displayed in Figure~\ref{ceva}.

\begin{theorem} {\bf (Ceva's theorem and its converse).}
Let $A,B,C$ be the vertices of a triangle and $D,E,F$ be three points on
the edges of the triangle opposite to $A,B,C$ respectively. Then, the 
lines passing through the points $(A,D)$, $(B,E)$ and $(C,F)$ are concurrent
if and only if
\bela{E4.14}
  \frac{\overline{AF}}{\overline{FB}}\, 
  \frac{\overline{BD}}{\overline{DC}}\,
  \frac{\overline{CE}}{\overline{EA}} 
  = 1,
\ela
where $\overline{PQ}/\overline{QR}$ denotes the ratio of directed lengths
associated with any three collinear points $P,Q,R$.
\end{theorem}
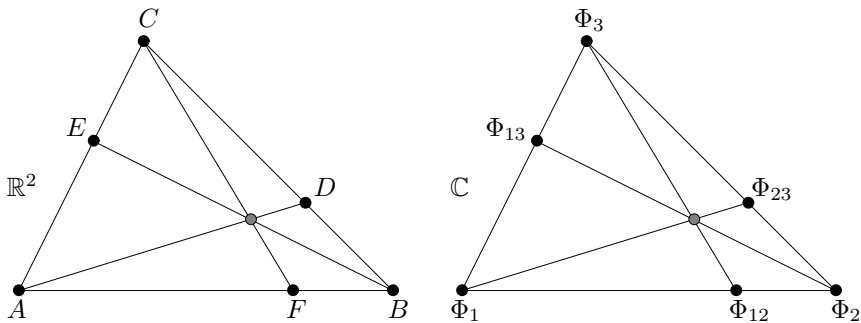
\begin{figure}[h]
\begin{center}
\setlength{\unitlength}{0.00057489in}
\begin{picture}(7543,2670)(0,-30)
\put(90,225){\blacken\ellipse{100}{100}}
\put(765,1575){\blacken\ellipse{100}{100}}
\put(2565,225){\blacken\ellipse{100}{100}}
\put(3465,225){\blacken\ellipse{100}{100}}
\put(1215,2475){\blacken\ellipse{100}{100}}
\put(2675,1015){\blacken\ellipse{100}{100}}
\path(90,225)(1215,2475)
\path(90,225)(3465,225)
\path(1215,2475)(3465,225)
\path(1215,2475)(2565,225)
\path(765,1575)(3465,225)
\path(90,225)(2691,1025)
\put(2185,865){\shade\ellipse{100}{100}}
\put(-20,1080){$\mathbb{R}^2$}
\put(-20,-20){$A$}
\put(3420,-20){$B$}
\put(1170,2600){$C$}
\put(2750,1080){$D$}
\put(510,1620){$E$}
\put(2500,-20){$F$}
\put(4090,225){\blacken\ellipse{100}{100}}
\put(4765,1575){\blacken\ellipse{100}{100}}
\put(6565,225){\blacken\ellipse{100}{100}}
\put(7465,225){\blacken\ellipse{100}{100}}
\put(5215,2475){\blacken\ellipse{100}{100}}
\put(6675,1015){\blacken\ellipse{100}{100}}
\path(4090,225)(5215,2475)
\path(4090,225)(7465,225)
\path(5215,2475)(7465,225)
\path(5215,2475)(6565,225)
\path(4765,1575)(7465,225)
\path(4090,225)(6691,1025)
\put(6185,865){\shade\ellipse{100}{100}}
\put(3980,1080){$\mathbb{C}$}
\put(3980,-20){$\Phi_1$}
\put(7420,-20){$\Phi_2$}
\put(5100,2600){$\Phi_3$}
\put(6700,1080){$\Phi_{23}$}
\put(4300,1620){$\Phi_{13}$}
\put(6500,-20){$\Phi_{12}$}
\end{picture}
\end{center}
\caption{A Ceva figure}
\label{ceva}
\end{figure}
Indeed, if $\Phi_1,\Phi_{12},\Phi_2,\Phi_{23},\Phi_3,\Phi_{13}$ are six points
defining a Ceva figure then there exist real shape parameters 
$\alpha,\beta,\gamma$ such that the linear system (\ref{E4.4}) is satisfied and
\bela{E4.15}
  M(\Phi_1,\Phi_{12},\Phi_2,\Phi_{23},\Phi_3,\Phi_{13}) = 1\quad
  \Leftrightarrow\quad \alpha\beta\gamma=-1.
\ela
However, if we demand that the above Ceva figure be part of a lattice with
real shape parameters then the linear system (\ref{E4.6}) must be satisfied
for some real shape parameters $\alpha_3,\beta_1,\gamma_2$. Consequently, the
points $\Phi_{12},\Phi_{23},\Phi_{13},\Phi_{123}$ must be collinear. This is
in contradiction with the Ceva figure but consistent with the Menelaus figure.
Thus, the multi-ratio 
$M(\Phi_1,\Phi_{12},\Phi_2,\Phi_{23},\Phi_3,\Phi_{13})$ is required to be 
$-1$. 

It is interesting to note that the linear system (\ref{E4.4}),
written in the form
\bela{E4.16}
  \bear{rlrl}
  \Phi_{12} = &a\Phi_1 + a'\Phi_2,&\quad a + a' = & 1\as
  \Phi_{23} = &b\Phi_2 + b'\Phi_3,&\quad b + b' = & 1\as
  \Phi_{13} = &c\Phi_3 + c'\Phi_1,&\quad c + c' = &1,
  \ear
\ela
has been exploited in \cite{Ped70} to give an algebraic proof of Menelaus' 
theorem, that is $abc=-a'b'c'$. Moreover, since $a,\ldots,c'$ are real, we
may think of $\Phi$ as a real two-dimensional vector-valued function
satisfying (\ref{E4.16}) and therefore
\bela{E4.17}
  \bear{rlrl}
  \Phi_{12}-\Phi = &a(\Phi_1-\Phi) + a'(\Phi_2-\Phi)\as
  \Phi_{23}-\Phi = &b(\Phi_2-\Phi) + b'(\Phi_3-\Phi)\as
  \Phi_{13}-\Phi = &c(\Phi_3-\Phi) + c'(\Phi_1-\Phi).
  \ear
\ela
The latter constitutes the well-known discrete Darboux system 
\cite{BogKon95,Dol97} on the plane if the constraints (\ref{E4.16})$_{2,4,6}$
are ignored. Hence, Menelaus lattices constitute a particular reduction of
Darboux lattices on the plane. In fact, if $\Phi$ is regarded as a vector
in $\mathbb{R}^n$ then the system (\ref{E4.16}) is still compatible and
$\Phi$ defines a conjugate lattice \cite{BobSei99} since the (degenerate)
quadrilaterals
$<\Phi,\Phi_i,\Phi_k,\Phi_{ik}>,\,i\neq k$ are planar. Thus, the following
theorem obtains:

\begin{theorem} {\bf (Menelaus-Darboux lattices).} The conjugate lattice
$\Phi\in\mathbb{R}^n$ defined by the compatible linear triad
\bela{E4.18}
  \Phi_{ik}-\Phi = \frac{\phi_k}{\phi_k-\phi_i}(\Phi_i-\Phi) +
                   \frac{\phi_i}{\phi_i-\phi_k}(\Phi_k-\Phi),\quad i\neq k,
\ela
where $\phi$ is a real 
solution of the discrete dual KP wave function equation
(\ref{E2.4.5})$_2$, consists of (planar) Menelaus figures with vertices
$\Phi_1,\Phi_{12},\Phi_2,\Phi_{23},\Phi_3,\Phi_{13}$.
\end{theorem}

\section{KdV, Schramm and Boussinesq reductions}
\setcounter{equation}{0}

The continuous KP equation and its hierarchy admit 1+1-dimensional reductions 
to the Korteweg-de Vries (KdV) and Boussinesq (BQ) equations and their 
associated hierarchies. The Schwarzian analogues of 
these hierarchies and their integrable discretizations may be obtained from 
the (discrete) SKP hierarchy~\cite{KPKDVBQ}. Here,  we discuss the 
algebraic and geometric properties of two-dimensional reductions of the 
dSKP equation.

\subsection{The discrete Schwarzian KdV equation}

\begin{figure}[h]
\begin{center}
\setlength{\unitlength}{0.00053489in}
\begin{picture}(8525,3119)(0,-10)
\put(1763,1769){\ellipse{2250}{2250}}
\put(1763,1432){\ellipse{1574}{1574}}
\put(975,1207){\ellipse{1934}{1934}}
\put(2888,1319){\ellipse{2624}{2624}}
\put(6263,1432){\ellipse{1574}{1574}}
\put(6263,1769){\ellipse{2250}{2250}}
\put(7163,1657){\ellipse{2708}{2708}}
\put(5700,1544){\ellipse{2122}{2122}}
\put(1763,644){\blacken\ellipse{100}{100}}
\put(6263,644){\blacken\ellipse{100}{100}}
\put(1343,2099){\blacken\ellipse{100}{100}}
\put(1923,2199){\blacken\ellipse{100}{100}}
\put(2538,2574){\blacken\ellipse{100}{100}}
\put(1693,1849){\blacken\ellipse{100}{100}}
\put(698,2129){\blacken\ellipse{100}{100}}
\put(5478,2574){\blacken\ellipse{100}{100}}
\put(6118,2514){\blacken\ellipse{100}{100}}
\put(6558,2154){\blacken\ellipse{100}{100}}
\put(6538,2849){\blacken\ellipse{100}{100}}
\put(5893,2124){\blacken\ellipse{100}{100}}
\put(1703,1514){$\scriptstyle1$}
\put(1643,329){$\scriptstyle23$}
\put(1853,2294){$\scriptstyle3$}
\put(2468,2714){$\scriptstyle13$}
\put(1268,2249){$\scriptstyle2$}
\put(458,2189){$\scriptstyle12$}
\put(6173,389){$\scriptstyle23$}
\put(6068,2609){$\scriptstyle1$}
\put(6443,2969){$\scriptstyle3$}
\put(5303,2669){$\scriptstyle2$}
\put(5613,2174){$\scriptstyle13$}
\put(6623,2249){$\scriptstyle12$}
\end{picture}
\end{center}
\caption{Degenerate Menelaus configurations}
\label{touch}
\end{figure}
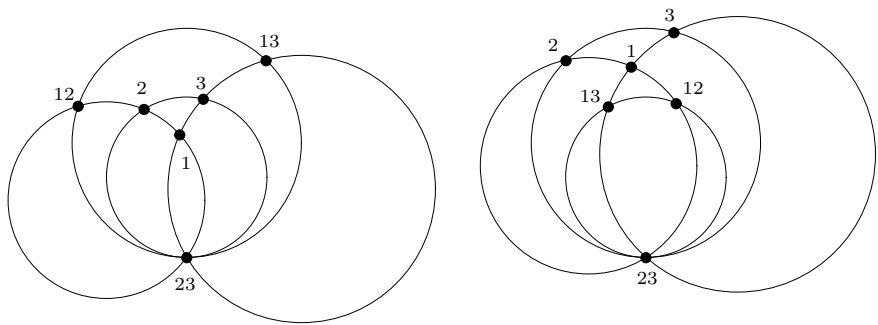

\begin{figure}[h]
\begin{center}
\setlength{\unitlength}{0.00052489in}
\begin{picture}(6324,2064)(0,-10)
\put(439,455){\blacken\ellipse{100}{100}}
\put(739,1137){\blacken\ellipse{100}{100}}
\put(919,1565){\blacken\ellipse{100}{100}}
\put(1362,1130){\blacken\ellipse{100}{100}}
\put(2029,455){\blacken\ellipse{100}{100}}
\put(4032,1587){\blacken\ellipse{100}{100}}
\put(4054,455){\blacken\ellipse{100}{100}}
\put(5187,1580){\blacken\ellipse{100}{100}}
\put(5659,455){\blacken\ellipse{100}{100}}
\put(4714,1107){\blacken\ellipse{100}{100}}
\path(1137,2037)(237,12)
\path(12,462)(2487,462)
\path(237,1137)(1812,1137)
\path(462,2037)(2487,12)
\path(3612,12)(5637,2037)
\path(3612,462)(6312,462)
\path(3387,2037)(6312,12)
\path(3387,1587)(5637,1587)
\put(462,232){$\scriptstyle12$}
\put(1902,232){$\scriptstyle13$}
\put(777,912){$\scriptstyle2$}
\put(867,1722){$\scriptstyle1$}
\put(3972,1697){$\scriptstyle13$}
\put(5052,1697){$\scriptstyle12$}
\put(4692,867){$\scriptstyle1$}
\put(4017,232){$\scriptstyle2$}
\put(5592,232){$\scriptstyle3$}
\put(1272,912){$\scriptstyle3$}
\end{picture}
\end{center}
\caption{Degenerate Menelaus figures}
\label{touchy}
\end{figure}
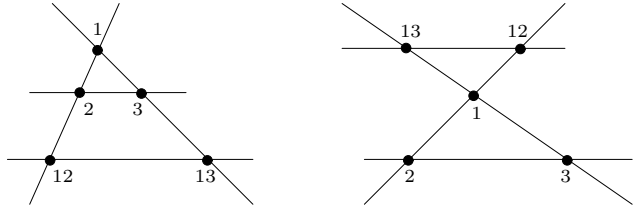

We first observe that the dSKP equation
$M(\Phi_1,\Phi_{12},\Phi_2,\Phi_{23},\Phi_3,\Phi_{13})
=-1$ may be written as equalities of cross-ratios. For instance, we may choose
the formulation
\bela{E5.4a}
  Q(\Phi_{23},\Phi_1,\Phi_{12},\Phi_2) = Q(\Phi_3,\Phi_{13},\Phi_1,\Phi_{23}),
\ela
where the cross-ratio $Q$ of four points $P_1,P_2,P_3,P_4$ on the complex plane
is defined~by
\bela{E5.3}
  Q(P_1,P_2,P_3,P_4) = \frac{(P_1-P_2)(P_3-P_4)}{(P_2-P_3)(P_3-P_4)}.
\ela
Thus, if we impose the constraint \cite{KPKDVBQ}
\bela{E5.1}
  \Phi_{23} = \Phi
\ela
then the dSKP equation 
reduces to
\bela{E5.2}
  \Delta_2 Q(\Phi,\Phi_1,\Phi_{12},\Phi_2) = 0,
\ela
whence
\bela{E5.4}
  Q(\Phi,\Phi_1,\Phi_{12},\Phi_2) = \nu(n_1),
\ela
where $\nu$ is an arbitrary function of $n_1$. If, in addition, we demand that
the function $\nu$ be preserved by the group of inversive transformations then 
$\nu$ must be real. This corresponds to a particular case of the discrete
Schwarzian KdV (dSKdV) equation \cite{NijCap95} and implies that the points
$\Phi,\Phi_1,\Phi_{12},\Phi_2$ lie on a circle. The dSKdV equation is therefore
located in the class of dSKP reductions of the type
\bela{E5.4b}
  Q(\Phi_{23},\Phi_1,\Phi_{12},\Phi_2) = Q(\Phi_3,\Phi_{13},\Phi_1,\Phi_{23})
  \in\mathbb{R}.
\ela
The latter are associated with degenerate Menelaus configurations with two
touching circles as displayed in Figure \ref{touch}.
These two circles
become parallel straight  
lines if an inversion with respect to $\Phi_{23}$ is applied.
The resulting configurations as depicted in Figure \ref{touchy} may be regarded
as degenerate Menelaus figures.
It is also noted that since $\nu$ is independent of $n_2$, the relation 
(\ref{E5.4}) defines
particular generalized discrete conformal mappings as introduced in 
\cite{BobPin96}. The case \mbox{$\mu = -1$} corresponds to the standard 
definition of discrete conformal mappings~\cite{BobPin96}. 
Accordingly, discrete conformal mappings give rise to special degenerate
Clifford lattices and their constituent Menelaus and $\mathcal{C}_4$ Clifford 
configurations. 

\subsection{\boldmath$P$ lattices and Schramm's circle patterns} 

\begin{figure}[h]
\begin{center}
\setlength{\unitlength}{0.00052489in}
\begin{picture}(7074,3082)(0,-10)
\put(1437,2370){\ellipse{1124}{1124}}
\put(1437,907){\ellipse{1800}{1800}}
\put(762,2145){\ellipse{1508}{1508}}
\put(2337,1357){\ellipse{2012}{2012}}
\put(1437,1807){\blacken\ellipse{100}{100}}
\put(4587,682){\blacken\ellipse{100}{100}}
\put(5262,2032){\blacken\ellipse{100}{100}}
\put(6612,2032){\blacken\ellipse{100}{100}}
\put(5937,682){\blacken\ellipse{100}{100}}
\put(687,1387){\blacken\ellipse{100}{100}}
\put(1099,2812){\blacken\ellipse{100}{100}}
\put(1999,2302){\blacken\ellipse{100}{100}}
\put(2157,360){\blacken\ellipse{100}{100}}
\path(4362,232)(5487,2482)
\path(5712,232)(6837,2482)
\path(4137,682)(6387,682)
\path(4812,2032)(7062,2032)
\put(672,1132){$\scriptstyle1$}
\put(2157,142){$\scriptstyle2$}
\put(987,2932){$\scriptstyle13$}
\put(2067,2442){$\scriptstyle23$}
\put(1482,1582){$\scriptstyle3$}
\put(1207,1897){$\scriptstyle12$}
\put(4587,412){$\scriptstyle1$}
\put(5937,412){$\scriptstyle2$}
\put(6657,1762){$\scriptstyle23$}
\put(5262,1762){$\scriptstyle13$}
\end{picture}
\end{center}
\caption{A 5-point Menelaus configuration and its image under inversion}
\label{schramm}
\end{figure}
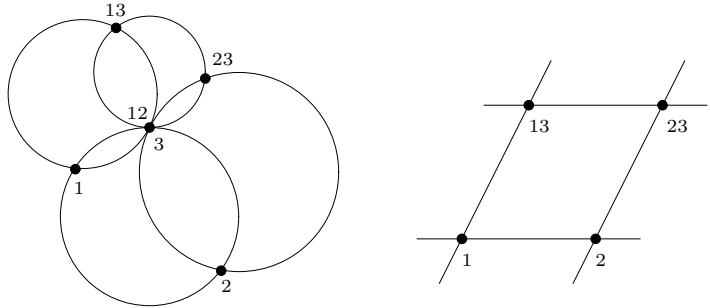

Further degeneration of the Menelaus configuration leads to two pairs of 
touching circles. Indeed, if two points with complementary indices coincide,
that is, for instance,
\bela{S1}
  \Phi_3 = \Phi_{12},
\ela
then a Menelaus configuration of the type displayed in Figure \ref{schramm}
is obtained. 
In fact, five points 
$\Phi_1,\Phi_{12},\Phi_2,\Phi_{23},\Phi_{13}$ on the complex plane
obey the multi-ratio condition
\bela{S2}
  M(\Phi_1,\Phi_{12},\Phi_2,\Phi_{23},\Phi_{12},\Phi_{13}) = -1
\ela
if and only if the pairs of `opposite' circles $S(\Phi_1,\Phi_{12},\Phi_2)$, 
$S(\Phi_{23},\Phi_{12},\Phi_{13})$ and $S(\Phi_{12},\Phi_2,\Phi_{23})$,
$S(\Phi_{12},\Phi_{13},\Phi_1)$ are tangent. In this case, inversion with 
respect to $\Phi_{12}$ leads to two pairs of parallel straight lines 
(see Figure \ref{schramm}). Thus, if we impose the condition (\ref{S1}) on a
Clifford lattice then the dSKP equation reduces to (\ref{S2}) or, equivalently,
\bela{S3}
  \frac{(\Phi_{\bar{2}}-\Phi)(\Phi_{\bar{1}}-\Phi_2)(\Phi-\Phi_1)}{
        (\Phi-\Phi_{\bar{1}})(\Phi_2-\Phi)(\Phi_1-\Phi_{\bar{2}})} = -1
\ela
and the geometry of the corresponding two-dimensional lattice is characterized
by the property that any vertex $\Phi$ and its neighbours 
$\Phi_1,\Phi_2,\Phi_{\bar{1}},\Phi_{\bar{2}}$ are positioned in such a way that
opposite circles through these points are tangent. Since inversion with respect
to any given vertex $\Phi$ produces a parallelogram, lattices of this kind
are termed $P$ lattices \cite{BobPin99}. It is known that the assumption that
the elementary quadrilaterals are inscribed in circles 
which intersect orthogonally is admissible and preserves 
integrability~\cite{BobPin99}. 
Moreover, if the elementary quadrilaterals are embedded then
Schramm's important circle patterns are retrieved~\cite{Sch97}. These may be
used to approximate analytic functions~(see~\cite{AgaBob00}).

\subsection{The discrete Schwarzian Boussinesq equation}

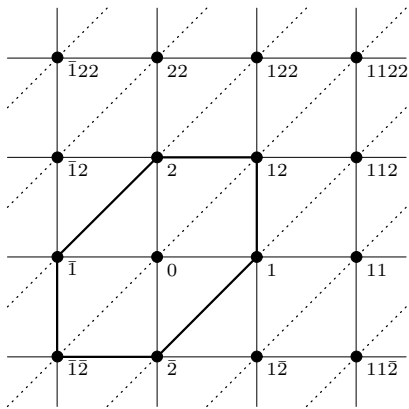
\begin{figure}
\begin{center}
\setlength{\unitlength}{0.00057489in}
\begin{picture}(3624,3639)(0,-10)
\put(462,3162){\blacken\ellipse{100}{100}}
\put(462,2262){\blacken\ellipse{100}{100}}
\put(462,1362){\blacken\ellipse{100}{100}}
\put(462,462){\blacken\ellipse{100}{100}}
\put(1362,3162){\blacken\ellipse{100}{100}}
\put(1362,2262){\blacken\ellipse{100}{100}}
\put(1362,1362){\blacken\ellipse{100}{100}}
\put(1362,462){\blacken\ellipse{100}{100}}
\put(2262,462){\blacken\ellipse{100}{100}}
\put(2262,1362){\blacken\ellipse{100}{100}}
\put(2262,2262){\blacken\ellipse{100}{100}}
\put(2262,3162){\blacken\ellipse{100}{100}}
\put(3162,3162){\blacken\ellipse{100}{100}}
\put(3162,2262){\blacken\ellipse{100}{100}}
\put(3162,1362){\blacken\ellipse{100}{100}}
\put(3162,462){\blacken\ellipse{100}{100}}
\path(12,1362)(3612,1362)
\path(12,462)(3612,462)
\path(462,3612)(462,12)
\path(1362,3612)(1362,12)
\path(2262,3612)(2262,12)
\path(3162,3612)(3162,12)
\dottedline{60.000}(12,1812)(1812,3612)
\dottedline{60.000}(12,12)(3612,3612)
\dottedline{60.000}(912,12)(3612,2712)
\dottedline{60.000}(1812,12)(3612,1812)
\dottedline{60.000}(2712,12)(3612,912)
\dottedline{60.000}(12,2712)(912,3612)
\dottedline{60.000}(12,912)(2712,3612)
\path(12,3162)(3612,3162)
\path(12,2262)(3612,2262)
\thicklines
\path(462,462)(462,1362)(1362,2262)
        (2262,2262)(2262,1362)(1362,462)(462,462)
\put(1452,1192){$\scriptstyle0$}
\put(2352,1192){$\scriptstyle1$}
\put(3252,1192){$\scriptstyle11$}
\put(1452,2092){$\scriptstyle2$}
\put(1452,2992){$\scriptstyle22$}
\put(2352,2092){$\scriptstyle12$}
\put(3252,2092){$\scriptstyle112$}
\put(2352,2992){$\scriptstyle122$}
\put(3252,2992){$\scriptstyle1122$}
\put(1452,292){$\scriptstyle\bar{2}$}
\put(552,1192){$\scriptstyle\bar{1}$}
\put(552,292){$\scriptstyle\bar{1}\bar{2}$}
\put(2352,292){$\scriptstyle1\bar{2}$}
\put(3252,292){$\scriptstyle11\bar{2}$}
\put(552,2092){$\scriptstyle\bar{1}2$}
\put(552,2992){$\scriptstyle\bar{1}22$}
\end{picture}
\end{center}
\caption{The dSBQ grid}
\label{boussinesq}
\end{figure}

The reduction \cite{KPKDVBQ}
\bela{E5.5}
  \Phi_{123} = \Phi
\ela
leads to the two-dimensional lattice equation
$M(\Phi_1,\Phi_{12},\Phi_2,\Phi_{\bar{1}},
\Phi_{\bar{1}\bar{2}},\Phi_{\bar{2}}) = -1$, that is
\bela{E5.6}
 \frac{(\Phi_1-\Phi_{12})(\Phi_2-\Phi_{\bar{1}})(\Phi_{\bar{1}\bar{2}}-
 \Phi_{\bar{2}})}{(\Phi_{12}-\Phi_2)(\Phi_{\bar{1}}-\Phi_{\bar{1}\bar{2}})
 (\Phi_{\bar{2}}-\Phi_1)} = -1,
\ela
which is known as the discrete Schwarzian Boussinesq (dSBQ) 
equation~\cite{Nij99}. dSBQ lattices are therefore defined by the property that
any sextuplet \linebreak $(\Phi_1,\Phi_{12},\Phi_2,\Phi_{\bar{1}},
\Phi_{\bar{1}\bar{2}},\Phi_{\bar{2}})$ represents a Menelaus 
configuration. Any six points of this kind constitute the vertices of a 
hexagon if we extend the square lattice to a triangular lattice by introducing 
one family of diagonals as indicated in Figure~\ref{boussinesq}. 
Since triangular lattices may be covered by three overlapping hexagonal 
sublattices, dSBQ lattices may also be interpreted as triangular lattices 
which are such that the three honeycomb lattices are composed of Menelaus 
configurations. These 
triangular lattices and their concomitant hexagonal lattices have recently been
introduced and discussed in detail in \cite{BobHofSur01} in the context of 
hexagonal circle patterns. The Menelaus connection presented 
here therefore provides a geometric interpretation of these lattices and, by
construction, their integrability is inherited from Clifford lattices. 
A geometric description of the multi-ratio condition (\ref{E5.6}) in the case 
of six concyclic points has been given earlier in \cite{BobHof01}.  

In conclusion, it is noted that there exist other canonical integrable 
reductions of the dSKP equation such as periodic fixed point reductions 
and symmetry reductions associated with the group of inversive transformations.
For instance, we may assume that
a Clifford lattice is invariant under inversion in a circle which is 
equivalent to considering Clifford lattices on a circle. Thus, if we
focus on the unit circle centred at the origin then the parametrization
\bela{E7.1}
   \Phi = e^{2\ii\omega}
\ela
reduces the dSKP equation to
\bela{E7.2}
  \frac{\sin(\omega_{1}-\omega_{12})\sin(\omega_{2}-\omega_{23})
        \sin(\omega_{3}-\omega_{13})}{\sin(\omega_{12}-\omega_{2})
        \sin(\omega_{23}-\omega_{3})\sin(\omega_{13}-\omega_{1})} = -1.
\ela
The corresponding `circular' lattices may be mapped to the complex plane
by means of a Combescure-type transformation which, in fact, exists for any
Clifford lattice. Indeed, it is readily verified that if $\Phi$ constitutes
a solution of the dSKP equation with corresponding shape parameters 
$\alpha,\beta,\gamma$ then the linear system
\bela{E7.3}
  \bear{rl}
   \rho_2-\rho = &\alpha(\rho_1-\rho)\as
   \rho_3-\rho = & \beta(\rho_2-\rho)\as
   \rho_1-\rho = &\gamma(\rho_3-\rho)
 \ear
\ela
is compatible. The compatibility of this linear system which may be regarded 
as adjoint to (\ref{E4.4}) guarantees that there exists a bilinear potential
$\Phi'$ defined~by
\bela{E7.4}
  \Phi'_i - \Phi' = \rho(\Phi_i-\Phi),\quad i=1,2,3.
\ela
It is readily verified that $\Phi'$ constitutes another solution of the
dSKP equation.

\section{The natural continuum limit. (Quasi-)confor-mal mappings}
\setcounter{equation}{0}

The standard continuum limit of the dSKP equation represented by
$\Phi_i(t) = \Phi(t+[a_i]),\, a_i\rightarrow0$ gives rise to the SKP
hierarchy \cite{BogKon98a,BogKon98b,BogKon99}. However, from a geometric
point of view, it is natural to consider the limit in which the
polygons $\Phi(n_i=\mbox{const},n_k=\mbox{const})$ become
coordinate lines on the complex plane. Thus, we regard the differences 
$\Delta_i\Phi = \Phi_i-\Phi$
between two neighbouring
points $\Phi_i$ and $\Phi$ as approximations of derivatives, that is
\bela{E6.1}
  \Phi_i = \Phi + \epsilon\Phi_{x_i} + O(\epsilon^2),\quad i=1,2,3,
\ela
where $\epsilon$ is a lattice parameter and $\Phi_{x_i}=\del\Phi/\del x_i$.
In the limit $\epsilon\rightarrow0$ and $(x_1,x_2,x_3)=(x,y,t)$, the dSKP 
then reduces to
\bela{E6.2}
  \left(\ln\frac{\Phi_{x}}{\Phi_{y}}\right)_{t} + 
  \left(\ln\frac{\Phi_{y}}{\Phi_{t}}\right)_{x} + 
  \left(\ln\frac{\Phi_{t}}{\Phi_{x}}\right)_{y} = 0,
\ela
while the linear system (\ref{E4.4}) reads
\bela{E6.3}
  \Phi_{y} = \alpha\Phi_{x},\quad
  \Phi_{t} =  \beta\Phi_{y},\quad
  \Phi_{x} = \gamma\Phi_{t}.
\ela
In addition to the group of inversive transformations, equation (\ref{E6.2}) 
is invariant under $\Phi\rightarrow F(\Phi)$, where $F$ is an arbitrary
differentiable function. The solutions of (\ref{E6.2}) constitute mappings
$\Phi : \mathbb{R}^3\rightarrow\mathbb{C}$ to which we shall refer as
SKP mappings. 

As in the discrete case, the linear system (\ref{E6.3}) implies
that the `continuous' shape factors $\alpha,\beta,\gamma$ are constrained by
$\alpha\beta\gamma=1$. Accordingly, if we set $\delta = 1/\gamma$ then
SKP mappings are obtained by integrating the linear pair
\bela{E6.4}
   \Phi_y = \alpha\Phi_x,\quad \Phi_t = \delta\Phi_x,
\ela
where $\alpha,\delta$ are solutions of the coupled nonlinear system
\bela{E6.5}
  \alpha_t + \alpha\delta_x = \delta_y + \delta\alpha_x,\quad
  (1-\delta)\alpha_t = (1-\alpha)\delta_y.
\ela
The latter comprises (\ref{E6.2}) written in terms of the shape factors 
and the compatibility condition for (\ref{E6.4}). An alternative form of
(\ref{E6.5}) is given by the single equation
\bela{E6.6}
  (e^{\varphi_y} - e^{\varphi_t})\varphi_{yt} +
  (1+e^{\varphi_t})\varphi_{xy} - (1+e^{\varphi_y})\varphi_{xt} = 0
\ela
with $\alpha = 1+e^{-\varphi_y},\,\delta=1+e^{-\varphi_t}$. In terms of the
variables 
\bela{E6.7}
  \mu = \frac{\i-\alpha}{\i+\alpha},\quad z = x+\i y, 
\ela
the linear system (\ref{E6.4}) reads
\bela{E6.8}
  \Phi_{\bar{z}} = \mu\Phi_z,\quad \Phi_t = \delta(\Phi_z+\Phi_{\bar{z}})
\ela
so that it becomes transparent that SKP mappings are descriptive of particular 
evolutions of quasi-conformal mappings if the variable $t$ is interpreted
as time. Here, we assume that the coefficient $\mu$ in the Beltrami equation
(\ref{E6.8})$_1$ is bounded. In other words, any SKP mapping consists of
a one-parameter family $\Phi(t)$ of quasi-conformal mappings.

\subsection{SKdV mappings}

The continuum limit of the constraint (\ref{E5.1}) corresponding to the dSKdV 
equation is readily seen to be
\bela{E6.9}
  \Phi_y + \Phi_t = 0.
\ela
Accordingly, $\beta = -1$ and $\alpha = \alpha(x)$. The associated
SKP mappings are then defined by the Beltrami equation 
\bela{E6.10}
  \Phi_{\bar{z}} = \mu(z+\bar{z})\Phi_z,
\ela
where $\mu$ is an arbitrary function of its argument. Thus, the SKdV mappings
$\Phi(t=\mbox{const}): \mathbb{C}\rightarrow \mathbb{C}$
constitute quasi-conformal mappings with complex dilations of the form
$\mu = \mu(z+\bar{z})$. If the shape factor $\alpha$ is purely
imaginary then we may assume without loss of generality that $\alpha=\pm \i$. 
This corresponds to $\mu=0$ or $\mu\rightarrow\infty$ and therefore
$\Phi_{\bar{z}}=0$ or $\Phi_z=0$. In this way, analytic or anti-analytic
functions are obtained. This is in agreement with the fact that the discrete
SKdV equation (\ref{E5.4}) with $\nu=-1$ reduces to $\Phi_x^2/\Phi_y^2=-1$ in
the continuum limit $\epsilon\rightarrow0$ and therefore defines
(anti-)analytic functions \cite{BobPin96}.

\subsection{SBQ mappings}

The continuum limit of the constraint (\ref{E5.5}) corresponding to the
dSBQ equation is given by
\bela{E6.11}
  \Phi_x + \Phi_y + \Phi_t= 0.
\ela
This implies that $1+\alpha+\alpha\beta=0$ and hence the associated
SKP mappings are obtained via
\bela{E6.12}
  \Phi_{\bar{z}} = \mu\Phi_z,\quad 
  \mu_{\bar{z}} = \frac{1-2\i\mu}{2\i+\mu}\mu_z.
\ela
Thus, the SBQ mappings $\Phi(t=\mbox{const}):\mathbb{C}\rightarrow\mathbb{C}$
constitute quasi-conformal mappings with complex dilations $\mu$ given 
implicitly by
\bela{E6.13}
  \mu = G((2\i+\mu)z + (1-2\i\mu)\bar{z}),
\ela
where $G$ is an arbitrary differentiable function.

To summarize, it has been shown that the scalar equation (\ref{E6.2}) and its
canonical reductions define classes of `integrable' quasi-conformal mappings.
The solutions of the dSKP equation and the dSKdV and dSBQ reductions 
(\ref{E5.4}), (\ref{E5.6}) may therefore be regarded as their integrable 
discretizations. 

\medskip
\noindent
{\bf Acknowledgement.} One of the authors (B.G.K.) is grateful to the
School of Mathematics, UNSW for the kind hospitality.

\end{document}